\documentclass[conference]{IEEEtran}

\usepackage[style=ieee,eprint=false,url=false,doi=true,backend=bibtex]{biblatex}
\usepackage{latexsym,upref,amsmath,amssymb,amsfonts,amsthm}
\usepackage{graphicx}

\usepackage{geometry}
\usepackage{hyperref}
\usepackage{xurl}

\usepackage{cleveref}
\Crefname{figure}{Fig.}{Figs.}
\Crefname{algorithm}{Alg.}{Algs.}

\usepackage{caption}
\usepackage{subcaption}
\usepackage{float}
\usepackage{enumerate}
\usepackage{enumitem}
\usepackage[usenames,dvipsnames,svgnames,table]{xcolor}
\usepackage{dsfont}

\usepackage{tikz}
\usepackage{pgfplots}
\pgfplotsset{compat=newest}
\usepgfplotslibrary{groupplots}
\usepgfplotslibrary{polar}
\usepgfplotslibrary{smithchart}
\usepgfplotslibrary{statistics}
\usepgfplotslibrary{dateplot}
\usepgfplotslibrary{ternary}
\usetikzlibrary{arrows.meta}
\usetikzlibrary{backgrounds}
\usepgfplotslibrary{patchplots}
\usepgfplotslibrary{fillbetween}
\pgfplotsset{%
layers/standard/.define layer set={%
    background,axis background,axis grid,axis ticks,axis lines,axis tick labels,pre main,main,axis descriptions,axis foreground%
}{grid style= {/pgfplots/on layer=axis grid},%
    tick style= {/pgfplots/on layer=axis ticks},%
    axis line style= {/pgfplots/on layer=axis lines},%
    label style= {/pgfplots/on layer=axis descriptions},%
    legend style= {/pgfplots/on layer=axis descriptions},%
    title style= {/pgfplots/on layer=axis descriptions},%
    colorbar style= {/pgfplots/on layer=axis descriptions},%
    ticklabel style= {/pgfplots/on layer=axis tick labels},%
    axis background@ style={/pgfplots/on layer=axis background},%
    3d box foreground style={/pgfplots/on layer=axis foreground},%
    },
}

\theoremstyle{plain}
\newtheorem{theorem}{Theorem}

\newtheorem{problem}{Problem}

\theoremstyle{definition}
\newtheorem{definition}[theorem]{Definition}

\def\G{\mathcal{G}}
\def\N{\mathcal{N}}
\def\E{\mathcal{E}}
\def\U{\underline{U}}
\def\B{\overline{B}}
\def\S{\mathcal{S}}
\def\C{\mathcal{C}}

\DeclareMathOperator*{\argmax}{arg\,max}
\DeclareMathOperator*{\argmin}{arg\,min}

\usepackage[ruled,section]{algorithm}
\usepackage{algpseudocode}
\algnewcommand{\IIf}[1]{\State\algorithmicif\ #1\ \algorithmicthen}
\algnewcommand{\EndIIf}{\unskip\ \algorithmicend\ \algorithmicif}
\algnewcommand{\FForAll}[1]{\State\algorithmicforall\ #1\ \algorithmicdo}
\algnewcommand{\EndFFor}{\unskip\ \algorithmicend\ \algorithmicfor}
\algblock[indent]{BeginIndent}{EndIndent}{}{}

\IEEEoverridecommandlockouts
\usepackage{amsmath,amssymb,amsfonts}
\usepackage{graphicx}
\usepackage{textcomp}
\usepackage{xcolor}
\begin{document}

\title{Algorithms for Interference Minimization in Future Wireless Network Decomposition
\thanks{Funding: PLE and TRM were supported in part by the National Research,
Development and Innovation Office -- NKFIH grant SNN~135643, K~132696.}
\thanks{T.\ R.\ Mezei and P.\ L.\ Erdős are affiliated with the Dept.\ of
Combinatorics and applications, Alfréd Rényi Inst.\ of Math.\ (LERN),
Budapest, Hungary (e-mail: \texttt{<erdos.peter, mezei.tamas.robert>@renyi.hu})}
\thanks{Y.\ Yu, X.\ Chen, W.\ Han, and B.\ Bai are with the Theory Lab, Huawei
    Technologies Co.\ Ltd.,\ Hong
Kong Science Park, Shatin, New Territories, Hong Kong (e-mail:
\texttt{<yu.yiding,  chenxiang73, harvey.hanwei, baibo8>@huawei.com}
)}
}

\author{
\IEEEauthorblockN{Péter L.\ Erdős, Tamás Róbert Mezei, Yiding Yu, Xiang Chen, Wei Han, and Bo Bai}
}

\maketitle

\begin{abstract}
We propose a simple and fast method for providing a high quality
solution for the sum-interference minimization problem. As future networks are deployed in high
density urban areas, improved clustering methods are needed to provide low
interference network connectivity. The proposed algorithm applies
straightforward similarity based clustering and optionally stable matchings to
outperform state of the art algorithms. The running times of our
algorithms are dominated by one matrix multiplication.
\end{abstract}

\section{Introduction}\label{sec:intro}
One of the typical problems in algorithmic graph theory is to assign the
vertices of a graph to partition clusters under some optimization condition.
These general problem formulations have ample applications in everyday life. One
of the very first applications of this kind was discussed by \citeauthor{KL70}
in \citeyear{KL70} (\cite{KL70}): let $G$ be an edge weighted graph (the weights
are real numbers). The goal is to partition the graph into classes of at most
$k$ vertices with minimum weighted edge cut. The graph itself represents a
complicated electronic design to be placed  on  printed circuit cards, where
each card can contain at most $k$ components and where the electronic
connections among the cards are expensive. They observed that there is very
small chance to solve the problem exactly. (The notion of NP-hardness was still
at least one year away.) Therefore they suggested a heuristic: let's start with
a feasible configuration, then improve the design by exchanging vertices between
the partition classes.

From that time on different graph partition problems are abundant in applied
graph theory. While the continuous clustering problems (for example, vertices on
surfaces or in higher dimensional spaces) can be often solved almost exactly
(for example, with the Lagrange multiplier method or the spectral clustering
methods, etc., see~\cite{Dhill04}), the graph partition problems usually need
sophisticated heuristics.  In \cite{Dhill07} the authors surveyed a
number of useful weight functions for the graph partitioning problem, and
studied extensions of vector based clustering methods to graph partitioning
problems.

Such graph partitioning problems arise naturally for large scale wireless
networks. Large scale cellular networks have been designed, from the beginning,
based on the idea of network decomposition~\cite{Tse05}. Dividing a large area
into a number of cells where a base station (BS) is placed at the center of each
cell and serves users who fall into its coverage. In this way the original
large-scale network is decomposed into multiple subnetworks which operate
independently. This idea is simple, but design is infected by the  well-known
cell-edge problem~\cite{Ges10}: users located at the cell edge area would suffer
from strong interference from the neighboring BSs.

With coordinated multipoint (CoMP) transmission from multiple BSs, the network
is decomposed into clusters of cells, where BSs in the same cluster would
jointly serve the users. %
Therefore the typical approach was to cluster the BSs, then distribute  the
users among the defined clusters. With a fixed and a-priori (the signal-strength
of the users are unknown) BS clustering pattern, nevertheless, users at the
cluster edge still suffer from strong interference from neighboring clusters.

In this paper we provide a simple and fast method for constructing clustering of
wireless networks with low sum-interference~\cite{Dhill07} that are applicable
for real-time clustering in future wireless networks. Our proposed algorithms are significantly
different from previous approaches and typically outperform those.

\subsection{Related Work}
\citeauthor{DaiBai} introduced a new network model (\cite{DaiBai}) for
optimizing the clustering process, where the BSs and the users are clustered in
parallel to minimize the sum-interference. \citeauthor{DaiBai} apply a widely
used eigenvector based method called spectral clustering to construct the
clusters. Spectral clustering solves a relaxed quadratic programming problem and
constructs the clusters by discretizing the continuous solution. This method
more-or-less represents the state of the art, so we will evaluate our methods in
comparison to it.

\section{Interference Minimization Problem}\label{sec:problem}

\subsection{Problem Description}

Let $\G=(V,\E)$ be the complete bipartite graph, with a non-negative real weight
function $w:\E\rightarrow \mathbb{R}^{\ge 0}$ on the edges.  One class, $U$,
denotes the users and the other class, $B$, denotes the set of all base stations
(BSs). (So $V=U \cup B$.) The weight function denotes the \emph{interference}
between a BS and a user, in the case if they are in different clusters under the
proposed vertex partition. (In the bipartite graph of course there is no edges
among users, and among BSs.)  Our goal is to minimize the total interference
(described later on) under the defined clustering. This description does not
ensure the condition that each user is served by some BSs. Therefore we add the
extrinsic condition, that there is no cluster which contains only users, but no
BSs. (Clusters with only BSs are allowed. The BSs in such clusters will be
turned off temporarily.)

For a vertex subset $S$ in the graph $G$ let $w(S)$ denote
\begin{equation}\label{eq1}
w(S):= \sum_{i,j\in S} w_{i,j} \quad \text{and}  \quad
\bar w(S):= \sum_{i\in S, j\not\in S} w_{i,j}.
\end{equation}
For a fixed the number $M\in \mathbb{N}$, take a partition $\C$ of $V$ with $M$
\emph{clusters}: $\C=\{C_1,\ldots,C_M\}$.
Consider the following optimization problem:
\begin{problem}[Interference minimization (\textbf{IM}) problem with fixed $M$]\label{P1}
\begin{equation}\label{eq2}
\min _{\C \in \Pi_M(\G)} \sum_{m=1}^M \frac{\bar w(C_m)}{w(C_m)}.
\end{equation}
\end{problem}
The objective function is not the sum of the \emph{cut numbers} but a normalized
one. The reason for this is that the purely cut number based optimization will
provide often very unbalanced partition class sizes. This is described in detail
in~\cite{DaiBai}, after equation~(3) of that paper.

\medskip

There is a technical problem hidden in \cref{eq2}: by the formulation of the
denominator, it can occur that a cluster contains zero BS or zero users. In both
cases the denominator is zero. If there is a cluster without BSs, then the users
of this cluster are not associated to any BS and the interference metric is
infinite (or undefined). When a cluster contains at least one BS and zero users,
then we can imagine that the stations will be switched off temporarily, that is,
they are not considered as a source of interference.

\medskip

To be able to use our procedure in practical application (like on-line
optimization of users' distribution among base stations in 5G mobile
networks) we have some further considerations. Our secondary  objectives are the
follows:
\begin{itemize}
    \item Phase 1: We want a fast, centralized algorithm to find an initial
        solution.
    \item Phase 2: During the calls the users may move away from the BSs of a
        given cluster, some may finish the calls, while others (currently not
        represented in the bipartite graph) may initiate calls. Therefore we
        need an incremental algorithm, that is able adaptively change the edge
        weights and/or can update the actual vertices. This phase must be
        initiated and managed  distributively by the users.
    \item In every few seconds Phase 1 should be executed again (centrally) to
        find a new optimal clustering solution.
\end{itemize}

In practical applications the clusters cannot be arbitrarily complex (from an
engineering point of view), therefore we consider an upper bound $T$ on the
possible numbers of the BSs in any cluster. This component is a new addition to
the model, it was not considered earlier. In previous work, the engineering
complexity of the BSs' clusters was handled indirectly. One possible way to do
so was suggested by \citeauthor{DaiBai} in~\cite{DaiBai}. 

At first they proved that the minimum value in~\eqref{eq2} is monotone
increasing as the value $M$ is increasing.
\begin{theorem}[\citeauthor{DaiBai}, Theorem~1 in~\cite{DaiBai}]\label{th:DB}
\begin{equation}
\min _{\C \in \Pi_M(\G)} \sum_{m=1}^M \frac{\bar w(C_m)}{w(C_m)} \le \min _{\C \in
\Pi_{M+1}(\G)} \sum_{m=1}^{M+1} \frac{\bar w(C_m)}{w(C_m)}.
\end{equation}
\end{theorem}
Then they introduced the \textbf{Max-Num} problem: here they want to increase
the number $M$ as long as the minimum value in \eqref{eq2} is still smaller than
a relative small, given positive number. \citeauthor{DaiBai} proposed a new approach to
solve this latter problem. At first they reformulated the question, using matrix
computation, to describe the constrains. This reformulation of the
\textbf{Max-Num} problem is NP-hard, due to the discretization. Then this was
relaxed to continuous constrains. Next a good heuristic was developed for the
problem, using a generalized spectral clustering method. Unfortunately, the
computational complexity of the method is still quite high for fast, practical
application. Furthermore the indirect approach does not ensure always that the
provided clusters are ``simple'' enough.

In the remaining part of this paper we propose a new heuristic to solve
Problem~\ref{P1} directly to overstep the previous weaknesses.

\section{Dot-Product Hierarchical Clustering for IM}
In this section we describe a new and simple heuristic for the \textbf{IM}
problem. We cluster BSs based on a new similarity measure. The novelty lies in
the fact that the clustering is made on the basis of a relation between BSs
which is derived from the relation among BSs and users. At first we discuss the
original problem formulation: the value $M$ is an input parameter. We will come back
later to the variation of  the problem where an upper bound is given on the
maximum size of BS clusters.

\subsection{Similarity Measure}
A cursory study of \cref{eq2} says that we want to decompose the graph in such
a way that each cluster contains high weight edges, while the cuts among the
clusters consist of low weight edges. Let's assume that  the weight function is
given via the matrix $w$ where the rows correspond to the BSs, and the columns
correspond to the users.  Then each $w_{i,j}$ is the weight between BS $i$ and
user $j$.

Let $w_{i,\bullet}$ denote the row of BS $i$, and let
$w_{\bullet,j}$ denote the column of user $j$. So $w =
{[w_{i,\bullet}]}_{i\in B}={[w_{\bullet,j}]}_{j\in U}$.  Our heuristics would
say that the larger the number of high weighted common neighbors of two
BSs, the more advantageous it is for the two BSs to be
included in the same cluster. So define the \emph{similarity function}
\begin{equation}\label{eq:dot}
	\rho : B \times B \rightarrow \mathbb{R}^{\ge 0} \quad \text{with} \quad
	\rho (i,j) := \frac{w_{i,\bullet}^T \cdot
	w_{j,\bullet}}{\|w_{i,\bullet}\|\cdot \|w_{j,\bullet}\|}
\end{equation}
among the BSs, where $\|\cdot\|$ is the Euclidean-norm. The enumerator
of \cref{eq:dot} is what we refer to by \emph{dot product}. The similarity
$\rho$ depends only on the weights between the users and the BSs.
Clearly, the bigger the product, the greater the similarity between the
BSs.

\medskip

In the interference minimization model, a set of BSs in a cluster behave as one BS. Indeed, if $\{C_1,\ldots,C_M\}\in \Pi_M(\mathcal{G})$ minimizes \cref{eq2}, then replacing the set of BSs in cluster $C_i$ with just one new BS $b_\text{new}$ whose weight to user $j$ is $\sum_{k\in C_i\cap B}w_{k,j}$ preserves the optimum, and the interference
metric takes this optimum on $\left\{C_1,\ldots,C_{i-1},C_i-B+\{b_\text{new}\},C_{i+1},\ldots,C_M\right\}$. Define
\begin{equation}\label{eq:vec}
    \mathrm{vec}(B_k)=\sum_{i\in B_k}w_{i,\bullet}
\end{equation}
as the sum of the signal strength vectors of the BSs in $B_k$. The similarity function $\rho$ can be naturally extended to sets of BSs:
\begin{equation}\label{eq:dot2}
    \begin{split}
        &\rho : 2^B \times 2^B \rightarrow \mathbb{R}^{\ge 0} \quad \text{with}\\
        &\rho (B_k,B_m) := \frac{{\mathrm{vec}(B_k)}^T \cdot
            \mathrm{vec}(B_m)}{\|\mathrm{vec}(B_k)\|\cdot \|\mathrm{vec}(B_m)\|}.
    \end{split}
\end{equation}

\subsection{Hierarchical Clustering: Defining BS Clusters}
Next we describe our hierarchical clustering algorithm: we call it
\textbf{DPH-clustering}, short for \textbf{dot-product hierarchical
clustering}. Let the fixed integer $M$ be the desired number of clusters.

There seems to be no clear leading method to cluster $B$ based on $\rho$. Our
choice is a simple \emph{hierarchical clustering} method: merge two clusters
that have the highest similarity $\rho$ between them until the desired number of
clusters is reached. As we will soon see, this works reasonably well. Here we
want to emphasize that using normalization in \cref{eq:dot,eq:dot2} is a natural
idea.

\medskip

Let the initial partition be $\mathcal{B}_0$ which contains a cluster for each
BS in $B$ (thus $|B|=| \mathcal{B}_0|$). We merge two clusters in each
of the $|B|-M$ rounds iteratively to obtain a sequence of partitions
$\mathcal{B}_0,\mathcal{B}_1, \ldots, \mathcal{B}_{|B|-M}$ of $B$, where
$|\mathcal{B}_r|=|B|-r$.  $\mathcal{B}_r$ is obtained from $\mathcal{B}_{r-1}$
by merging the two clusters of $\mathcal{B}_r$ with the largest similarity
$\rho$ between them as defined by \cref{eq:dot2}.

In \Cref{alg:clusteringB} we will maintain $\rho(B_k,B_m)$ for every $B_k,B_M\in
\mathcal{B}_r$ as follows. Let us
define the symmetric function $\mathit{dot}$ for every $k=1,\ldots,M$ as
\begin{align}\label{eq:dotsum}
\mathit{dot}(B_k,B_m)={\mathrm{vec}(B_k)}^T \cdot \mathrm{vec}(B_m).
\end{align}
If $\mathit{dot}$ is already computed for every pair in $\mathcal{B}_r\times \mathcal{B}_r$,
then $\rho$ can be computed via three scalar operations for any pair
of clusters in $\mathcal{B}_r\times \mathcal{B}_r$, since
\begin{align}\label{eq:rhodot}
    \rho(B_k,B_m)=\frac{\mathit{dot}(B_k,B_m)}{\sqrt{\mathit{dot}(B_k,B_k)\cdot\mathit{dot}(B_m,B_m)}}.
\end{align}

\begin{algorithm}
	\caption{Hierarchical clustering based on the similarity function $\rho$}\label{alg:clusteringB}
	\begin{algorithmic}
		\Function{DPH-clustering}{$B,w,M$}
        \State $\mathcal{B}_0\gets \binom{B}{1}$
        \State $\mathit{dot}\gets w\cdot w^T$ \Comment{matrix multiplication}
		\For{$r=0$ to $|B|-M-1$}
        \State $\displaystyle\{B',B''\}\gets\argmax_{\{B_k,B_m\}\in
            \binom{\mathcal{B}_r}{2}}
            \rho(B_k,B_m)$\Comment{\eqref{eq:rhodot}}
        \State $\mathcal{B}_{r+1}=\mathcal{B}_r-\{B',B''\}+\{B'\cup
        B''\}$%
        \State update $\mathit{dot}$ \Comment{$\mathcal{O}(b)$ scalar operations}
		\EndFor
        \State\Return $\mathcal{B}_{|B|-M}$
		\EndFunction
	\end{algorithmic}
\end{algorithm}

The running time of \Cref{alg:clusteringB} is easily seen to be in
$\mathcal{O}\left(b^2u+(b-M)\cdot b^2\right)$, because when two clusters are
merged, $\mathit{dot}$ can be updated by summing the corresponding two rows and
two columns. Moreover, we may store the $\rho$-values of pairs in
$\mathcal{B}_r\times \mathcal{B}_r$ in a max-heap: when two clusters are merged,
at most $2b$ values need to be removed and at most $b$ new values need to be
inserted into the heap which contains the at most $\binom{b}{2}$ elements of the
set $\{\rho(B_k,B_m)\ |\ \{B_k,B_m\}\in \binom{\mathcal{B}_r}{2}\}$. With these
optimizations, the \textbf{for}-loop takes at most $\mathcal{O}\left((b-M)\cdot
b\log b\right)$ steps, thus the running-time of the algorithm is dominated by
the matrix multiplication $w\cdot w^T$. There are many techniques to accelerate
the multiplication of  matrices, which we do not discuss here, but let us
mention that if $b\simeq u$ then $w$ can be padded with zeros to a square
matrix, whose multiplication can be tackled with recursive divide-and-conquer
methods. We will discuss alternative strategies for the case when $b$ is much
larger than $u$ in \Cref{sec:Ularge}.

\subsection{Hierarchical Clustering: Assigning Users to BS Clusters}
Let $\mathcal{B}_{|B|-M}=\{B_1,\ldots,B_M\}$ be the final partition produced by
the hierarchical clustering. The final output will be of the
form $\mathcal{C}=\cup_{k=1}^M\{ B_i\cup U_i \}$, so it only remains to find a
clustering $\mathcal{U}=\{U_1,\ldots,U_M\}$ of $U$. We assign each user $j\in U$
to the cluster $U_\ell$ where
\begin{equation}\label{eq:user}
	\ell=\argmax_{k\in [1,M]} \sum_{i\in B_k}w_{i,j}.
\end{equation}

\begin{algorithm}
    \caption{Dot-product hierarchical clustering on $B$ then assigning each
    element of $U$ to the best cluster. The algorithm is relatively efficient if
$|U|$ is not much smaller than $|B|$.}\label{alg:dpBbestU}
	\begin{algorithmic}
        \Function{Similarity Clustering}{$B,U,w,M$}
        \State $B_1,\ldots,B_M\gets \Call{DPH-clustering}{B,w,M}$
        \State $U_1,\ldots,U_M\gets $ empty clusters
        \ForAll{$j\in U$}
	        \State $\ell\gets\argmax_{k\in [1,M]} \sum_{i\in B_k}w_{i,j}$
            \State add $j$ to $U_\ell$
        \EndFor
        \State\Return$\{ B_k\cup U_k\ |\ U_k\neq\emptyset\}$
		\EndFunction
	\end{algorithmic}
\end{algorithm}

The assignment defined by \cref{eq:user} is easy to compute, and it is trivial
to assign new users to a cluster. It may happen that a BS cluster is
left without users: such clusters are discarded at cost of decreasing the number
of clusters, thus the output clustering will not reach the target cardinality
$M$.

Discarding clusters that only intersect one of the classes is not an issue if
the hierarchical clustering is performed on $B$. However, were we to call
$\textsc{DPH-clustering}(U,w^T,M)$ to take advantage of computing a
smaller matrix product, we might discard user-clusters without BSs. To avoid
creating clusters without BSs, we supply alternative Phase~2 algorithms for
assigning elements of the yet unclustered class to the clusters of the already
DPH-clustered class in \Cref{sec:Ularge}.

\medskip

Since we had chosen the spectral clustering method as our baseline, next we
analyse the differences between the two approaches: \Cref{alg:dpBbestU} has
immediate advantages over the spectral clustering method:
\begin{enumerate}
	\item Hierarchical clustering is much faster than the spectral
		clustering method: the running time is dominated by multiplying two
        matrices of size $b\cdot u$.
    \item Using $\rho$ as the similarity function, the slight movements of the
        users change the similarity measure only slightly, therefore we may
        assume that the BS clustering is not necessarily updated in
        real time; a periodic (every couple hundred milliseconds) updating of
        $\mathcal{B}$ will be sufficient. This seems to be an adequate answer
        for the problem of Phase~2.
    \item It is easy to modify the hierarchical clustering to respect an upper
        bound $T$ on the size of the BS clusters, see
        \Cref{sec:further}.
\end{enumerate}

This concludes the description of our method in the case when there are fewer
BSs than users. In practice we expect this to be the case. However,
\Cref{alg:dpBbestU} does not perform efficiently in simulations
(see \Cref{sec:experiments}) if the number of BSs is far fewer than the number of
users. We deal with this case in the following section.

\section{When There  Are More BSs Than Users}\label{sec:Ularge}
Suppose that $B$ has many more elements than $U$. We can switch the roles of $B$
and $U$, and perform the hierarchical clustering (\Cref{alg:clusteringB}) on $U$
instead of $B$. There is a large computational advantage over the original
approach, since the running time is dominated by the complexity of the
matrix multiplication of $w^T\cdot w$ vs.\ $w\cdot w^T$. However, the slight
asymmetry in evaluating \cref{eq2} that we hinted at earlier becomes dangerous:
given a clustering $\mathcal{U}=\{U_1,\ldots,U_M\}$, if we assign each BS
$i\in B$ to $B_\ell$ (i.e., cluster $C_\ell$) where
\begin{equation}\label{eq:bs}
    \ell=\argmax_{k\in [1,M]} \sum_{j\in U_k}w_{i,j},
\end{equation}
we may end up with $C_k=U_k$ for some $k$. Let us describe two possible
solutions to avoid BS-less clusters.

\subsection{Assigning a BS to User Clusters Via Maximum Cardinality Matchings}
One way to overstep (not to solve) this problem is simply assigning a unique
BS to each user-cluster. After that we can assign the remaining
BSs in whatever manner we chose, for example, as described by
\cref{eq:bs}. Choosing these unique BS is not necessarily a trivial
task, it is equivalent to finding a matching of $\mathcal{U}$ into $B$ where the
edges have relatively large weights, preferably.

We try to find a maximum cardinality (and maximum weight) matching of
$\mathcal{U}$ into $B$ such that if $i\in B$ is matched to $U_k$ then $w(U_k\cup
\{i\})\neq 0$ (so that the interference cannot be infinite, no matter how we
complete the clustering). Even if the complexity of the maximum cardinality
matching is prohibitive in some of our applications, there exist approximate
solutions that provide a log-linear complexity. Using the algorithm of
\citeauthor{Duan}~\cite{Duan}, the $(1-\varepsilon)$-approximate solution for the matching
can be computed in $\mathcal{O}(bu\varepsilon^{-1}\log\varepsilon^{-1})$ time.
Therefore the running time of \Cref{alg:dpUbestB} is also dominated by the
matrix multiplication $w^T\cdot w$ in \textsc{DPH-clustering}.

\begin{algorithm}
    \caption{Dot-product hierarchical clustering on $U$, then assigning one BS
    to each cluster of $U$, and assigning the remaining BSs to the best
    cluster. The algorithm is relatively efficient if $|B|$ is larger than
    $|U|$.}\label{alg:dpUbestB}
	\begin{algorithmic}
        \Function{DPH+matching+best}{$B,U,w,M$}
        \State $U_1,\ldots,U_M\gets \Call{DPH-clustering}{U,w^T,M}$
        \State $B_1,\ldots,B_M\gets $ empty clusters
        \State $E\gets$ approx.\ max.\ card.\ max.\ weight matching $\{U_1,\ldots,U_M\}$
        into $B$ using positive weight edges, see~\cite{Duan}
        \ForAll{$i\in B$}
        \If{$U_\ell$ is joined to $i$ in $E$}
            \State add $i$ to $B_\ell$
            \Else
	        \State $\ell\gets\argmax_{k\in [1,M]} \sum_{j\in U_k}w_{i,j}$
            \State add $i$ to $B_\ell$
            \EndIf
        \EndFor
        \State\Return$\{ B_k\cup U_k\ |\ k=1,\ldots,M\}$
		\EndFunction
	\end{algorithmic}
\end{algorithm}

\subsection{Clustering \texorpdfstring{$B$}{B} via Stable Matchings}\label{sec:stable}
In this subsection we provide an alternative method based on the \emph{stable
matching} approach. A matching between two classes of entities is stable if
there is no pair of entities that both prefer each other over their current
match. The problem to find such a matching has many applications in economics,
see the works of Roth and Shapley~\cite{nobelprize2012}. In a generalization of
this problem entities in the first class can be matched to many entities of the
second class; this version is colloquially known as the college admissions
problem (many-to-one matching).

The stable matching algorithm of Gale and Shapley~\cite{GS62} can be used to
overcome the base-station-less cluster problem. In our (many-to-one) stable
matching setup, each cluster in $\mathcal{U}$ and each BS in $B$ is assigned a
list of real numbers corresponding to the members of the other class.  We are
looking for a (many-to-one matching) which is stable with respect to the
preference values: if $U_1b_1$ and $U_2b_2$ are in the stable matching, then we
must have
\begin{align*}
    \mathrm{pref}(b_1;U_1)&\ge \mathrm{pref}(b_1;U_2)\text{ or}\\
    \mathrm{pref}(U_2;b_2)&\ge \mathrm{pref}(U_2;b_1).
\end{align*}
Let the preference of BS $i\in B$ for the user-cluster $U_k$ be
\begin{equation}\label{eq:pref1}
    {\mathrm{pref}}(i;U_k)=\sum_{j\in U_k}w_{i,j},
\end{equation}
that is, the first preference of a base-stations $i$ is the user cluster
assigned by \cref{eq:bs}. However, we define the preferences asymmetrically. Let
the preference of cluster $U_k$ for BS $i\in B$ be
\begin{equation}\label{eq:pref2}
    {\mathrm{pref}}(U_k;i)=-\frac{\sum_{j\in U}w_{i,j}}{\sum_{j\in
    U_k}w_{i,j}}.
\end{equation}
Note the negative sign in \cref{eq:pref2}: the cluster prefers a small fraction
in absolute value. The reasoning for these preference values will be explained
shortly.

The stable matching algorithm can be used to find not just one matching, but a
complete clustering of the not yet clustered class $B$ (stable marriage vs.\
college admissions). This is equivalent to finding a many-to-one matching of $B$
to $\mathcal{U}$. Given a clustering
$\mathcal{U}=\{U_1,\ldots, U_M\}$ of $U$ (constructed by, say,
\textsc{DPH-clustering}), we set the capacity and usage of $B_k$ as follows:
\begin{align}\label{eq:capacity}
    \mathrm{capacity}(k)&=\sum_{j\in U_k}\sum_{i\in B}w_{i,j},\\
    \mathrm{usage}(k)&=\sum_{i\in B_k}\sum_{j\in U}w_{i,j}.
\end{align}
A BS can be added (matched) to $B_k$ without extra maintenance steps as long as
$\mathrm{usage}(k)\le\mathrm{capacity}(k)$ holds even after the BS
joins $B_k$. If
$\mathrm{usage}(k)>\mathrm{capacity}(k)$ after a BS joins $B_k$, then
remove the lowest-preference BS from $B_k$ if and only if
$\mathrm{usage}(k)\ge\mathrm{capacity}(k)$ holds even after removal.

\begin{definition}[Stable clustering]
A clustering $B_1,\ldots,B_M$ of $B$ is \textbf{stable} if
\begin{itemize}
    \item $\mathrm{usage}(k)-\sum_{j\in U}w_{\beta_k,j}\ge\mathrm{capacity}(k)$ where
$\beta_k=\argmin_{\lambda\in B_k}\mathrm{pref}(U_k;\lambda)$ for every $k=1,\ldots,M$, and
    \item for every $k\neq m$ and $\beta\in B_k$, $\gamma\in B_m$ we have
            $\mathrm{pref}(\beta;U_k)\ge \mathrm{pref}(\beta;U_m)$ or
            $\mathrm{pref}(U_m;\gamma)\ge \mathrm{pref}(U_m;\beta)$.
\end{itemize}
\end{definition}

\begin{algorithm}
    \caption{Extending a clustering $\mathcal{U}$ via stable matchings}\label{alg:stableB}
	\begin{algorithmic}
        \Function{Stable clustering}{$B,U,w,M$}
        \State$U_1,\ldots,U_M\gets \Call{DPH-clustering}{U,w^T,M}$
        \State$B_1,\ldots,B_M\gets$ empty clusters
        \While{$\exists i\in B\setminus B_1\setminus\ldots\setminus B_M$}
        \State let $U_k$ maximize $\mathrm{pref}(i;U_k)$ among
         clusters which have not rejected $i$
        \State add $i$ to $B_k$
        \State let $\beta\gets\argmin_{\lambda\in B_k}\mathrm{pref}(U_k;\lambda)$
        \While{$\mathrm{usage}(k)-\sum_{j\in U}w_{\beta,j}\ge\mathrm{capacity}(k)$}
        \State remove $\beta$ from $B_k$, i.e., $U_k$ rejects $\beta$
        \State let $\beta\gets\argmin_{\lambda\in B_k}\mathrm{pref}(U_k;\lambda)$
        \EndWhile
        \EndWhile
        \State\Return$\{ B_k\cup U_k\ |\ k=1,\ldots,M\}$
		\EndFunction
	\end{algorithmic}
\end{algorithm}

If $\mathrm{usage}(k)\ge\mathrm{capacity}(k)$ holds at some point during the
execution of \Cref{alg:stableB}, then it holds at any later step too.  Thus if a
BS $\beta$ is rejected by each cluster, then the usage of every
cluster increased above its capacity even without $\beta$'s contribution. This
is a contradiction by the handshaking lemma, since
$\sum_{k=1}^M\mathrm{capacity}(k)=\sum_{i\in B}\sum_{j\in U}w_{i,j}$. Therefore
\Cref{alg:stableB} terminates after at most $bu$ cycles of the outer
\textbf{while}-loop, and when it terminates, every BS is associated to
a cluster.

In other words, if $B_1,\ldots,B_M$ is a clustering of $B$, then the sum of
$\mathrm{usage}(k)$ is equal to the sum of $\mathrm{capacity}(k)$, which means that we can expect that there exists a small $\varepsilon>0$
such that for every $k$ we have $\mathrm{capacity}(k)\le
(1+\varepsilon)\cdot\mathrm{usage}(k)$. If that is so, then let $-c_k=\min_{i\in
B_k}{\mathrm{pref}}(U_k;i)$; we have
\begin{align*}
    \frac{\bar{w}(C_k)}{w(C_k)}&=\frac{\mathrm{capacity}(k)+\mathrm{usage}(k)}{w(C_k)}-2\le \\
    &\le
    (2+\varepsilon)\frac{\mathrm{usage}(k)}{w(C_k)}-2 \le (2+\varepsilon)c_k-2.
\end{align*}
Note that we have equality if the BSs in $B_k$ are equally preferred
by cluster $U_k$. In words, the higher the preferences of the associated
BSs, the lower the sum-interference is, so \cref{eq:pref2} is a
reasonable choice, because we can more or less guarantee
$\mathrm{capacity}(k)\simeq \mathrm{usage}(k)$ for each cluster $U_k$.

Observe, that the preferences can be extracted in $\mathcal{O}(Mb)$ time from
the computations performed by the hierarchical clustering of $U$.  By using
binary heaps to represent the clusters $B_1,\ldots,B_M$, we can insert and
remove BSs in $\log b$ time. Since every BS tries to join
each cluster at most once, the total running time is in $\mathcal{O}(Mb\log b)$.
This is clearly dominated by the complexity of matrix multiplication in the
hierarchical clustering algorithm.

\section{Experimentation}\label{sec:experiments}
We have compared the performance of Similarity~Clustering (\Cref{alg:dpBbestU}),
Stable~Clustering (\Cref{alg:stableB}), and Spectral~Clustering~\cite{DaiBai}
in several scenarios. In each case, base-stations (BSs) and users are placed
independently uniformly and randomly into $[0,1000]^2$ (a square with an area of
$1\,\mathrm{km}^2$). The weight (or signal
strength) between a BS and a user $b_i,u_j\in[0,1000]^2$ is set to
\begin{equation*}
    w_{i,j}=\left\{
        \begin{array}{ll}
            {\|\mathrm{dist}_{\min}\|}^{-\alpha} & \text{ if }
            \|b_i-u_j\|\le \mathrm{dist}_{\min},\\
            {\|b_i-u_j\|}^{-\alpha} & \text{ if }
            \mathrm{dist}_{\min}\le\|b_i-u_j\|\le \mathrm{dist}_{\max},\\
            0 & \text{ if }\mathrm{dist}_{\max}<\|b_i-u_j\|,
        \end{array}
        \right.
\end{equation*}
where we set the path loss component $\alpha=4$ (see~\cite{DaiBai,Tse05}),
$\mathrm{dist}_{\min}=1$, and $\mathrm{dist}_{\max}=200$.

For the unconstrained \Cref{P1} simulations show short running times and very
low interference measures on the acquired clusterings compared to the spectral clustering
method (at least when the number of users is larger than the number of BSs).
While the paper~\cite{DaiBai} suggests that we should expect a small number of
giant clusters, this is clearly not the case in our test runs. This phenomenon
requires much better understanding.

\begin{figure}
    \centering
    \begin{subfigure}{0.45\textwidth}
        \centering
        \includegraphics[width=\linewidth]{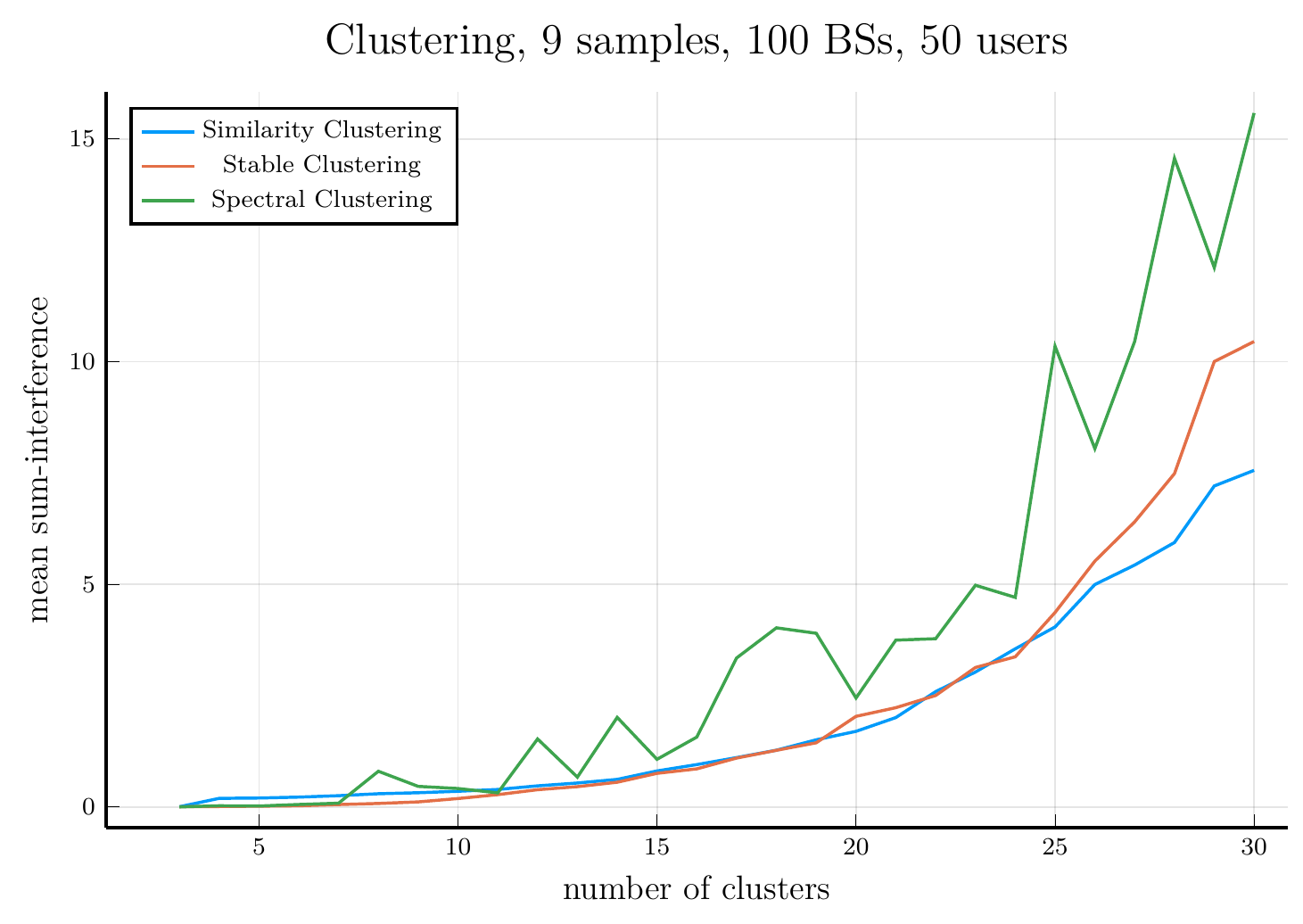}
        \caption{For $M\le 17$, Stable~Clustering is consistently better than
            Similarity~Clustering, which we suspect is another advantage of
            performing the DPH-clustering on the users' side}
    \end{subfigure}
    \begin{subfigure}{0.45\textwidth}
        \centering
        \includegraphics[width=\linewidth]{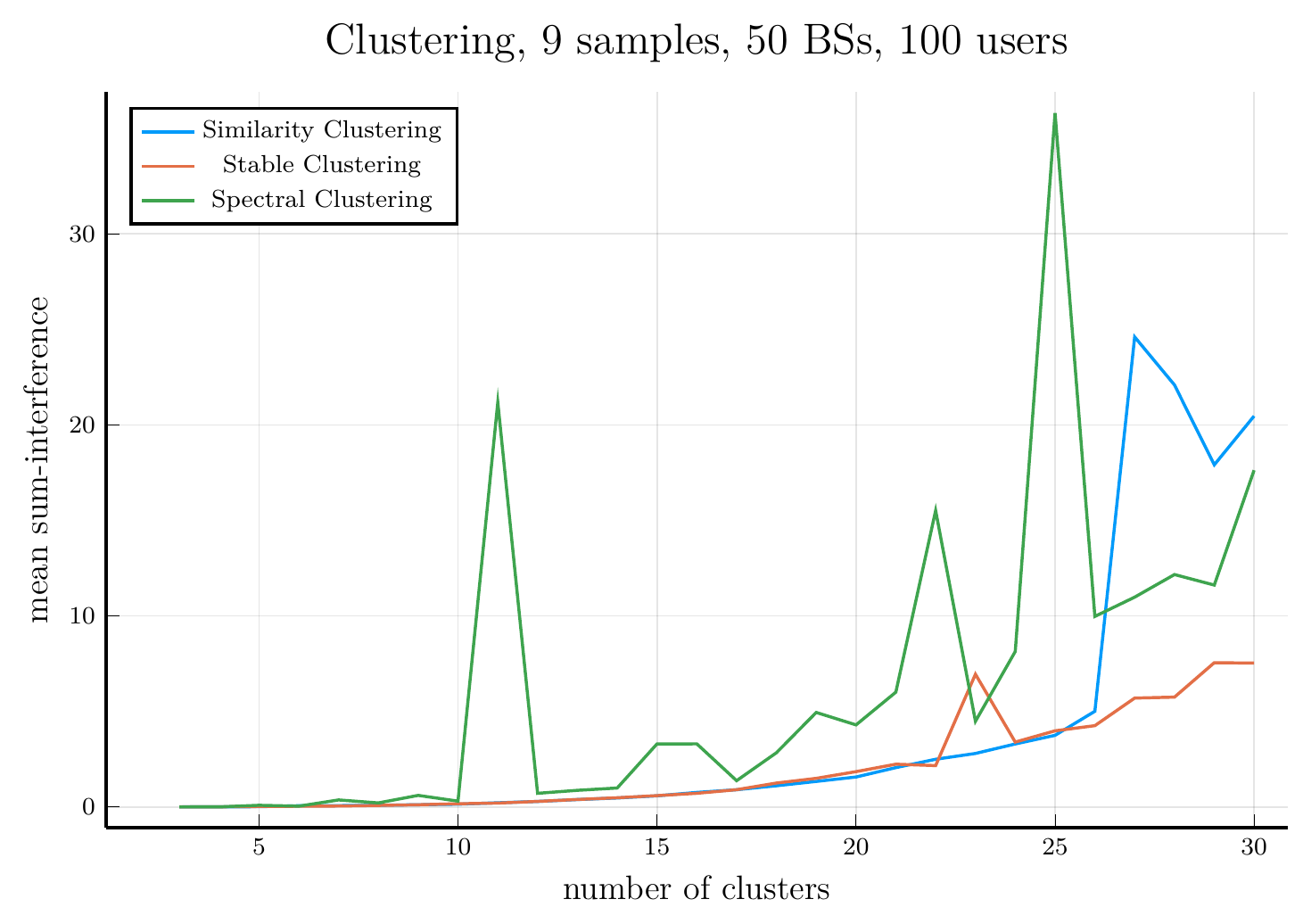}
        \caption{For $M\ge 27$ clusters, \Cref{alg:dpBbestU} produces some clusters
            with very weak BS coverage}
    \end{subfigure}
    \begin{subfigure}{0.45\textwidth}
        \centering
        \includegraphics[width=\linewidth]{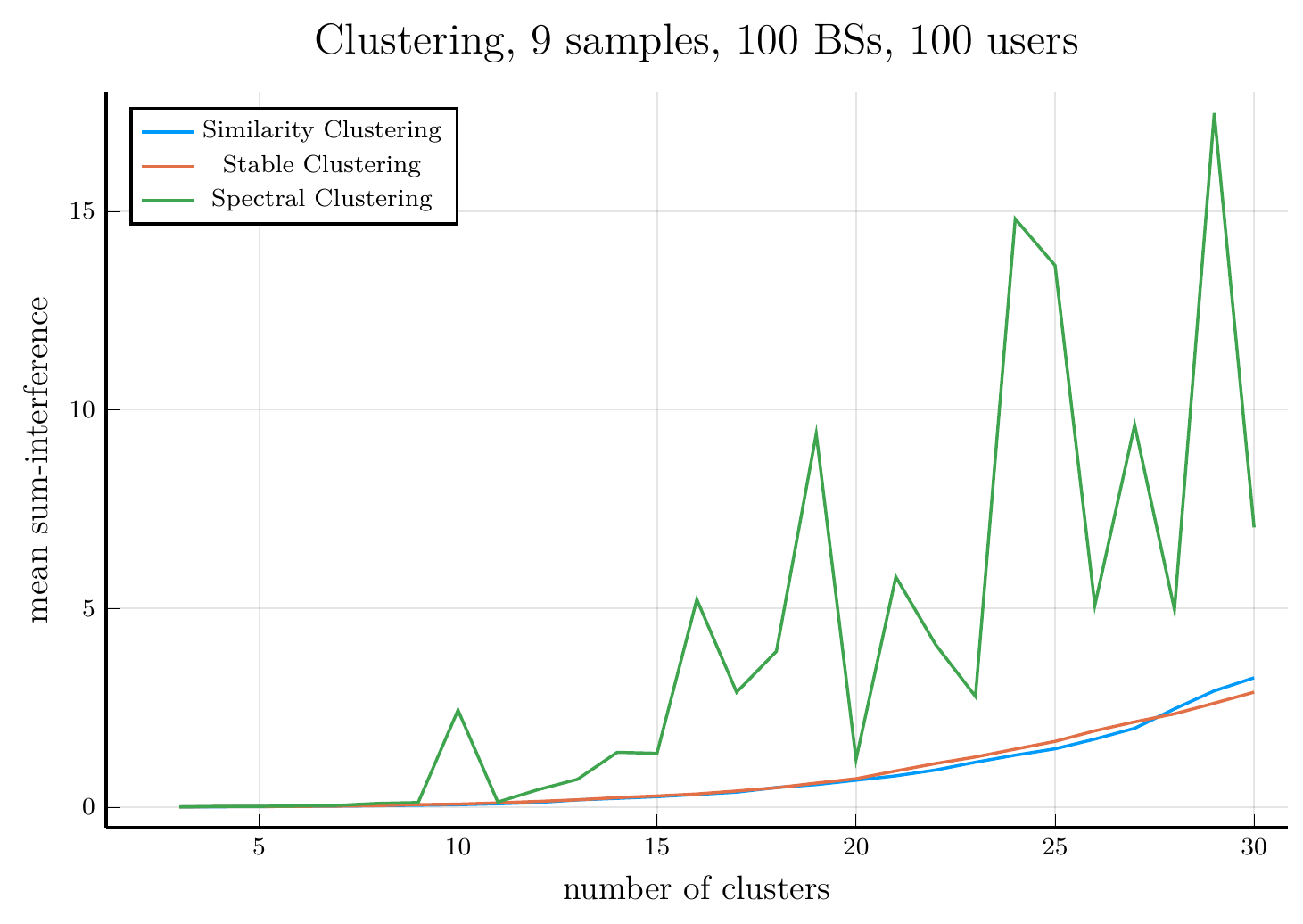}
        \caption{The performance of both of our algorithms scale
            much more evenly than the performance of the output of Spectral~Clustering}
    \end{subfigure}
    \caption{Mean sum-interference values as a function the number of clusters
    $M$}\label{fig:plots}
\end{figure}

\begin{figure}
    \centering
    \begin{subfigure}{0.49\textwidth}
        \centering
        \input{plots/SimilarityBestClusterBU_b100_u50_seed8_alpha4.0.tikz}
        \caption{Similarity Clustering, \Cref{alg:dpBbestU}}
    \end{subfigure}
    \begin{subfigure}{0.49\textwidth}
        \centering
        \input{plots/SimilarityPlusStableUB_b100_u50_seed8_alpha4.0.tikz}
        \caption{Stable Clustering, \Cref{alg:stableB}}
    \end{subfigure}
    \begin{subfigure}{0.49\textwidth}
        \centering
        \input{plots/SpectralClustering_b100_u50_seed8_alpha4.0.tikz}
        \caption{Spectral Clustering, see~\cite{DaiBai}}
    \end{subfigure}
    \caption{Comparison of the output of three clustering algorithms on 100
    BSs and 50 users, $M=10$. Triangles and circles represent BSs and users,
    respectively}\label{fig:comparison1}
\end{figure}

\begin{figure}
    \centering
    \begin{subfigure}{0.49\textwidth}
        \centering
        \input{plots/SimilarityBestClusterBU_b50_u100_seed4_alpha4.0.tikz}
        \caption{Similarity Clustering, \Cref{alg:dpBbestU}}
    \end{subfigure}
    \begin{subfigure}{0.49\textwidth}
        \centering
        \input{plots/SimilarityPlusStableUB_b50_u100_seed4_alpha4.0.tikz}
        \caption{Stable Clustering, \Cref{alg:stableB}}
    \end{subfigure}
    \begin{subfigure}{0.49\textwidth}
        \centering
        \input{plots/SpectralClustering_b50_u100_seed4_alpha4.0.tikz}
        \caption{Spectral Clustering, see~\cite{DaiBai}}
    \end{subfigure}
    \caption{Comparison of the output of three clustering algorithms on 50 BSs
    and 100 users, $M=10$. Triangles and circles represent BSs and users,
    respectively}\label{fig:comparison2}
\end{figure}

\Cref{fig:plots} compares the performance of the three mentioned algorithms in
three different settings: when there are many more BSs than users and vica
versa, and when there number of BSs is equal to the number of users.  The plots
correspond to the mean sum-interference values of the solutions provided by the
algorithms over 9 random samples of BSs-user placements. In all three cases we
find that our algorithms perform more consistently than the spectral clustering
method, and the performance of \Cref{alg:dpBbestU} and \Cref{alg:stableB} are
similar when the number of clusters are not too large or the number of BSs is
not much larger than the number of users.

\Cref{fig:comparison1} (50 BSs and 100 users) and \Cref{fig:comparison2} (100
BSs and 50 users) show example clusterings on the same BS-user placements each,
respectively.

\section{Conclusion and Further Considerations}\label{sec:further}
This paper proposed a similarity based hierarchical clustering method for
simple-and-fast wireless network decomposition in future wireless networks, with
the goal of minimizing sum interference in the overall network. Moreover, stable
matching were utilized to match BSs and users. Compared with state-of-the-art
spectrum clustering method, simulation results demonstrated that our proposed
algorithm could achieve better performance with much less complexity. Further
considerations are listed below, which lead to future research directions.
\begin{itemize}
    \item Suppose that the final clustering $\mathcal{B}$ on $B$ is restricted
        to clusters of size at most $T$. Running the DPH-clustering algorithm on
        $B$, we reject merging clusters whose total size is larger
        than $T$, i.e., we restrict our search for the largest similarity to
        pairs whose union has cardinality at most $T$. This allows one to
        control the maximum engineering complexity that arises in any cluster.
        The main issue with similarity clustering in this setting is that we run into
        discretization problems if $T$ is relatively small.
    \item It is also relatively easy to meaningfully modify \Cref{alg:stableB},
        to say, not assign a BS to a lower preference than half of
        their maximum. For example, we may specify that if $i\in B_k$ in the
        final clustering then
        \begin{equation*}
            \mathrm{pref}(i,U_k)\ge \frac12\max_{\lambda\in \{1,\ldots,M\}}\mathrm{pref}(i,U_\lambda).
        \end{equation*}
        If $i$ is rejected even by the least favored admissible cluster, then
        $i$ does not try to join clusters later on its preference list. Instead,
        when the Gale-Shapley algorithm completes, $i$ joins its most
        preferred cluster.
    \item Our proposed algorithms cluster every BS, even if using a BS in any
        cluster is causes more interference than not using it at all. This
        problem can be dealt with a trivial post-processing procedure: after the
        clusters $C_1,\ldots,C_M$ are determined, delete a tower $i$ from
        $C_k$ if doing so decreases $\overline{w}(C_k)/w(C_k)$.
\end{itemize}

\renewcommand*{\bibfont}{\small}
\printbibliography[]

\end{document}